\documentclass[10pt,letter,prl,twocolumn]{revtex4-1}

\usepackage{lineno}  

\usepackage{graphicx}  

\usepackage{xspace}
\usepackage{color}
\usepackage{colortbl}

\usepackage{amsmath}
\usepackage{tabularx}

\usepackage{ifthen} 

\newboolean{pdflatex}
\setboolean{pdflatex}{true} 
\newboolean{uprightparticles}
\setboolean{uprightparticles}{false} 
\usepackage{amssymb}
\usepackage{amsfonts}
\usepackage{upgreek}

\usepackage{lhcb-symbols-def}

\newboolean{articletitles}
\setboolean{articletitles}{true}
\usepackage{mciteplus}

\usepackage{hyperref}
\hypersetup{colorlinks=false,pdfborderstyle={/S/U/W 1}}
\usepackage[all]{hypcap} 

\begin{document}




\begin{titlepage}
\pagenumbering{roman}

\centerline{\Large EUROPEAN ORGANIZATION FOR NUCLEAR RESEARCH (CERN)}
\vspace*{1.0cm}
\hspace*{-0.5cm}
\begin{flushright}
  CERN-PH-EP-2012-050 \\  
  LHCb-PAPER-2011-028 \\  
  \today \\ 
\end{flushright}
\begin{picture}(0,0)
  \put(0,-30){\includegraphics[width=.14\textwidth]{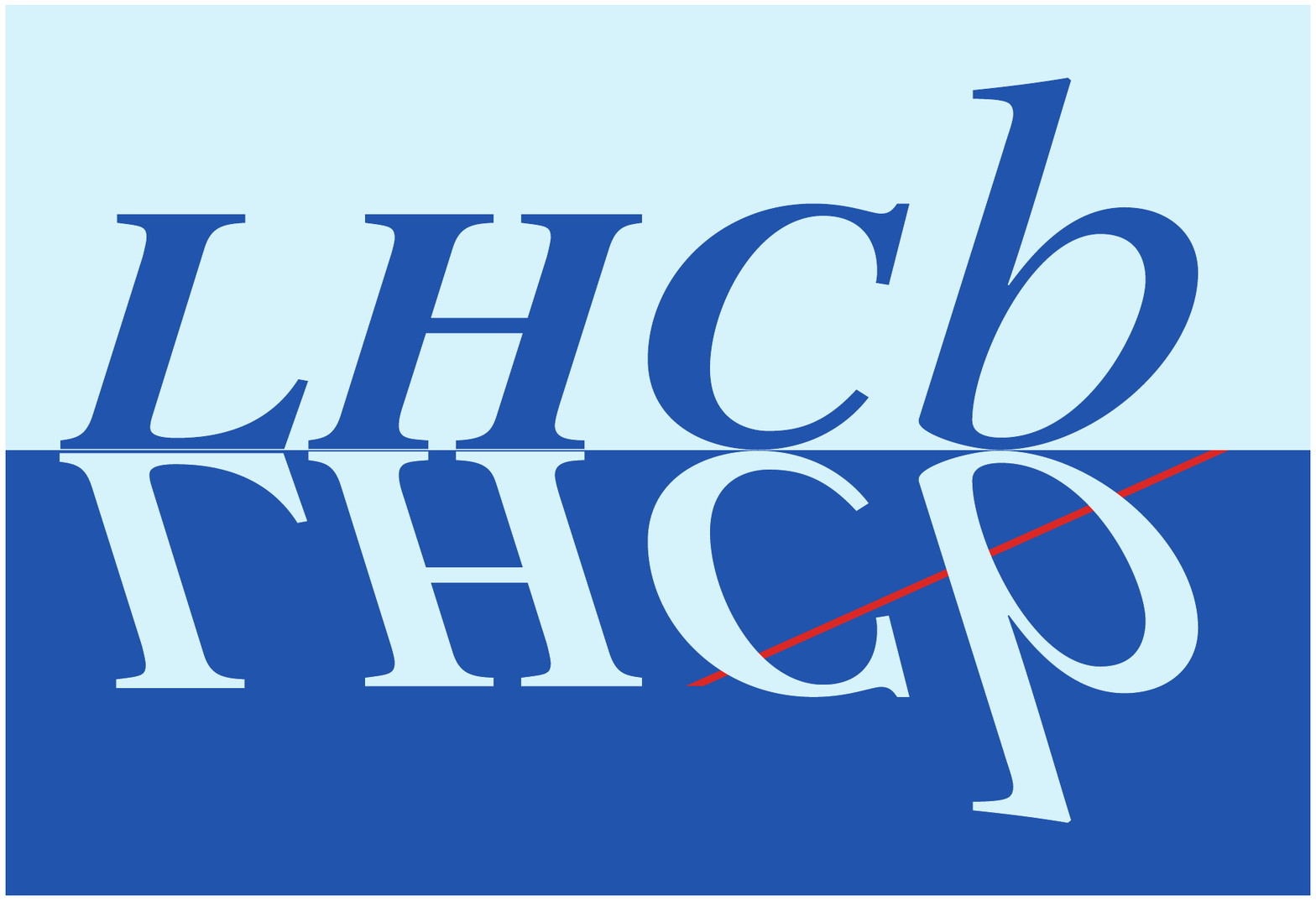}}
\end{picture}

\vspace*{1.0cm}

{\bf\boldmath\huge
\begin{center}
Determination of the sign of the decay width difference  in the \Bs\ system 
\end{center}
}

\vspace*{1.0cm}

\begin{center}
The LHCb collaboration
\end{center}

\vspace*{5mm}

\begin{flushleft}
R.~Aaij$^{38}$, 
C.~Abellan~Beteta$^{33,n}$, 
B.~Adeva$^{34}$, 
M.~Adinolfi$^{43}$, 
C.~Adrover$^{6}$, 
A.~Affolder$^{49}$, 
Z.~Ajaltouni$^{5}$, 
J.~Albrecht$^{35}$, 
F.~Alessio$^{35}$, 
M.~Alexander$^{48}$
% ...

{\footnotesize \it
% ...
$ ^{5}$Clermont Universit\'{e}, Universit\'{e} Blaise Pascal, CNRS/IN2P3, LPC, Clermont-Ferrand, France\\
$ ^{6}$CPPM, Aix-Marseille Universit\'{e}, CNRS/IN2P3, Marseille, France\\
% ...
$ ^{33}$Universitat de Barcelona, Barcelona, Spain\\
$ ^{34}$Universidad de Santiago de Compostela, Santiago de Compostela, Spain\\
$ ^{35}$European Organization for Nuclear Research (CERN), Geneva, Switzerland\\
% ...
$ ^{38}$Nikhef National Institute for Subatomic Physics, Amsterdam, The Netherlands\\
% ...
$ ^{43}$H.H. Wills Physics Laboratory, University of Bristol, Bristol, United Kingdom\\
% ...
$ ^{48}$School of Physics and Astronomy, University of Glasgow, Glasgow, United Kingdom\\
$ ^{49}$Oliver Lodge Laboratory, University of Liverpool, Liverpool, United Kingdom\\
% ...
$ ^{n}$LIFAELS, La Salle, Universitat Ramon Llull, Barcelona, Spain\\
% ...
}
\bigskip
\end{flushleft}

\cleardoublepage


\begin{abstract}

The interference between the $K^+K^-$ S-wave and P-wave amplitudes in 
$B_s^0 \rightarrow  J/\psi K^+K^-$  decays with the $K^+K^-$ pairs in the region around the $\phi(1020)$ resonance
is used to determine the variation of the difference of the strong phase between these amplitudes as a 
function of $K^+K^-$ invariant mass. Combined with the results from our $\CP$ asymmetry measurement in 
 $B_s^0 \rightarrow J/\psi \phi$ decays, we conclude that the $B_s^0$ mass eigenstate 
that is almost $\CP =+1$ is lighter and decays faster than the mass eigenstate that is almost $\CP =-1$. 
This determines the sign of the decay width difference $\DGs\equiv \GL -\GH$ to be positive.
Our result  also resolves the ambiguity in the past measurements of 
 the $\CP$ violating phase $\phis$ to be close to zero rather than $\pi$. 
These conclusions are in agreement with the Standard Model expectations.

  \bigskip
  \begin{center}
    Published on  Physical Review Letters
  \end{center}

\end{abstract}

\maketitle


\end{titlepage}





\pagestyle{plain} 
\setcounter{page}{1}
\pagenumbering{arabic}


The decay time distributions of $\Bs$ mesons decaying into the $J/\psi \phi$ final state have been used to measure the 
  parameters $\phis$ and $\DGs\equiv \GL -\GH$ of the \Bs\ system \cite{Abazov:2011ry, CDF:2011af, LHCb:2011aa}.
 Here  $\phis$ is the  $\CP$ violating phase equal to 
the phase difference between the amplitude for the direct \mbox{decay} and the amplitude for the decay after oscillation. 
$\GL$ and   $\GH$ are the decay widths of  the light  and heavy  $B_s^0$ mass 
eigenstates, respectively.
The most precise results, presented recently by the LHCb experiment \cite{LHCb:2011aa}, 
\begin{equation}
\begin{array}{lcl}
\phis & =& 0.15\phantom{0} \pm 0.18\phantom{0}\, ({\rm stat}) \pm 0.06\phantom{0} \, {\rm ( syst)~rad}, \\
\DGs &=& 0.123 \pm 0.029\, ({\rm stat}) \pm 0.011 {\rm \, (syst)~ps}^{-1},
\end{array} 
\end{equation}
show no evidence of $\CP$ violation yet, indicating that $\CP$ violation is rather small in the $B^0_s$ system. 
There is  clear evidence for the decay width difference  $\DGs$ being non-zero. It must be noted that there exists another solution
\begin{equation}
\begin{array}{lcl}
\phis & =&\phantom{-}2.99\phantom{0} \pm 0.18\phantom{0}\, ({\rm stat}) \pm 0.06\phantom{0} \,  {\rm ( syst)~rad}, \\
\DGs &=& -0.123 \pm 0.029 \, ({\rm stat}) \pm 0.011 \, {\rm (syst)~ps}^{-1},
\end{array}
\end{equation}
arising from the fact that the time dependent differential decay rates are invariant under the 
transformation $(\phi_{s},~\Delta \Gamma_{s}) \leftrightarrow (\pi - \phi_{s},~-\Delta \Gamma_{s})$ 
together with an appropriate transformation for the strong phases. 
In the absence of $\CP$ violation, $\sin \phi_s=0$, i.e. $\phi_s=0$ or $\phi_s=\pi$, 
the two mass eigenstates also become $\CP$ eigenstates with $\CP =+1$ and $\CP =-1$,
according to the relationship between \Bs\ mass eigenstates and \CP\ eigenstates given in Ref.~\cite{Dunietz:2000cr}.
They can be identified by the decays into \mbox{final} states which are $\CP$ eigenstates. 
In  $B_s^0 \rightarrow J/\psi  K^+ K^-$ decays,
the final state is a superposition of $\CP =+1$ and $\CP =-1$ for the $K^+ K^-$ pair in the P-wave configuration
and $\CP =-1$ for the $K^+ K^-$ pair in the S-wave configuration.
Higher order partial waves are neglected.
These decays have different angular distributions of the final state particles
 and are  distinguishable. 

Solution~I is close to the case $\phi_s=0$ and leads to the light (heavy) mass eigenstate being  almost 
aligned with the $\CP =+1$ $(\CP =-1)$ state. 
Similarly, solution~II is close to the case $\phi_s=\pi$ and leads to the heavy (light)  mass eigenstate 
being almost aligned with the $\CP =+1$ $(\CP =-1)$ state.  
In Fig. 2 of Ref.~\cite{LHCb:2011aa},
a fit to the observed decay time distribution shows that it can be well described by a superposition of two exponential functions
corresponding to $\CP =+1$ and $\CP =-1$, compatible with no $\CP$ violation~\cite{LHCb:2011aa}. 
In this fit the lifetime of the decay to the $\CP =+1$ final state is found to be smaller than that of the decay to $\CP =-1$. 
Thus the mass eigenstate that is predominantly $\CP$ even decays faster than the $\CP$ odd state.
For solution~I, we find $\DGs >0$, i.e. $ \GL > \GH$,  and for solution~II,  $\DGs <0$, i.e. $\GL<\GH$. 
In order to determine if  the decay width difference \DGs\  is positive or negative,
it is necessary to resolve the ambiguity between the two solutions.

Since each solution corresponds to a different set of strong phases, one may attempt to resolve the ambiguity by 
using the strong phases either as predicted by factorisation or as measured in $B^0  \rightarrow J/\psi K^{*0}$ decays. 
Unfortunately these two possibilities lead to opposite \mbox{answers}~\cite{Nandi:2008rg}. A direct experimental resolution of the ambiguity is 
therefore 
desirable.

In this Letter, we resolve this ambiguity using
 the \mbox{decay} $\BsToJPsiKK$ with $\Jpsi \rightarrow\mu^+ \mu^-$. 
The total decay amplitude is a coherent sum of S-wave and P-wave contributions.
The phase of the P-wave  amplitude, which can be described by a spin-1 Breit-Wigner function
of the invariant mass of the \KpKm\ pair, denoted by $m_{KK}$,
 rises  rapidly through the \particle{\phi (1020)} mass region.
On the other hand, the phase of the S-wave amplitude should vary relatively slowly for either  an $f_0 (980)$ contribution or
a nonresonant contribution. As a result, the phase difference between the S-wave and P-wave amplitudes falls rapidly
 with increasing   $m_{KK}$.
By measuring this phase difference   as a function of $m_{KK}$
and  taking the solution  with a decreasing trend around the  \particle{\phi (1020)} mass
as the physical solution, the sign of \DGs\ is determined and 
the ambiguity in \phis\ is resolved~\cite{Xie:2009fs}.
This is similar to the  way  the BaBar collaboration measured the
sign of $\cos2\beta$ using the decay  $B^0 \to J/\psi \KS \pi^0$~\cite{Aubert:2004cp},
where $2\beta$ is the weak phase  characterizing mixing-induced $\CP$ asymmetry
in this decay.

\begin{figure}[htbp]
  \centerline{
     \includegraphics[angle=0,width=0.5\textwidth]{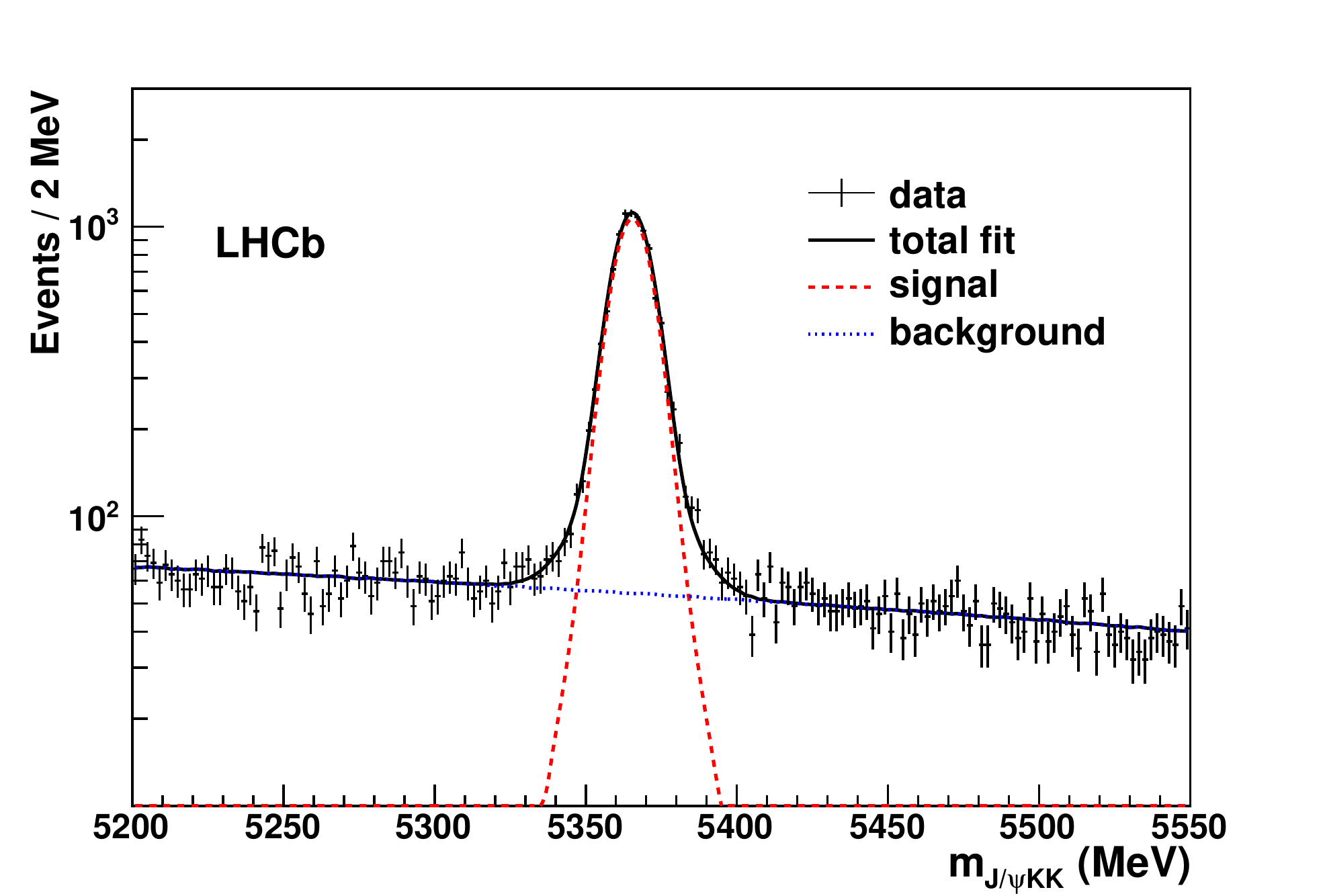}
  }
  \caption{ 
Invariant mass distribution for $\Bs\to \mumu K^+K^-$
    candidates, with the mass of the $\mu^+\mu^-$ pair  constrained
    to the nominal $J/\psi$ mass. The result of the fit is shown with
    signal (dashed curve) and  combinatorial background (dotted curve)
    components     and their sum (solid curve).  
}
\label{fig:mb}
\end{figure}

The analysis is based on the same data sample as used in Ref.~\cite{LHCb:2011aa},
which corresponds  to an integrated luminosity of 0.37 \invfb\ of $pp$ collisions collected
 by the LHCb experiment at the Large Hadron Collider
 at the centre of mass energy of $\sqrt{s}=7\TeV$.
 The LHCb detector is a forward spectrometer and is described in detail in Ref.~\cite{Alves:2008zz}.
The trigger, event selection criteria and analysis method are  very similar to those 
 in Ref.~\cite{LHCb:2011aa}, and here we  discuss only the differences.
The fraction of  \KpKm\ S-wave contribution measured
within $\pm$12 MeV  of  the nominal \particle{\phi (1020)} mass is $0.042 \pm 0.015 \pm 0.018$~\cite{LHCb:2011aa}.
(We adopt units such that $c=1$ and $\hbar=1$.)
The S-wave fraction depends on the mass range taken around the \particle{\phi (1020)}.
The result of Ref.~\cite{LHCb:2011aa} is consistent with the CDF limit on the S-wave fraction of less than $6\%$ at $95\%$ CL 
(in the range 1009--1028 MeV)~\cite{CDF:2011af}, smaller than
the \Dzero{} result of $(12\pm 3)\%$ (in 1010--1030 MeV)~\cite{Abazov:2011hv} and consistent with phenomenological 
expectations~\cite{Stone:2008ak}. 
In order to apply the ambiguity resolution method described above,  the range of $m_{KK}$ is extended  to 988--1050 MeV.
Figure~\ref{fig:mb} shows the $\mumu \KpKm $ mass distribution
where the mass of the $\mumu$ pair is constrained to the nominal \Jpsi\ mass.
We perform an unbinned maximum likelihood fit to the invariant mass distribution of the selected
\Bs\ candidates. 
The probability density function (PDF) for the signal \Bs\ invariant mass $m_{J/\psi KK}$ is modelled by
two Gaussian functions with a common mean.   
The fraction of the wide Gaussian and its width relative to that of 
the narrow Gaussian are fixed to values obtained from simulated events. 
A linear function describes the $m_{J/\psi KK}$ distribution of the background, which is dominated by combinatorial background.

\begin{figure}[htbp]
  \centerline{
    \includegraphics[angle=0,width=0.5\textwidth]{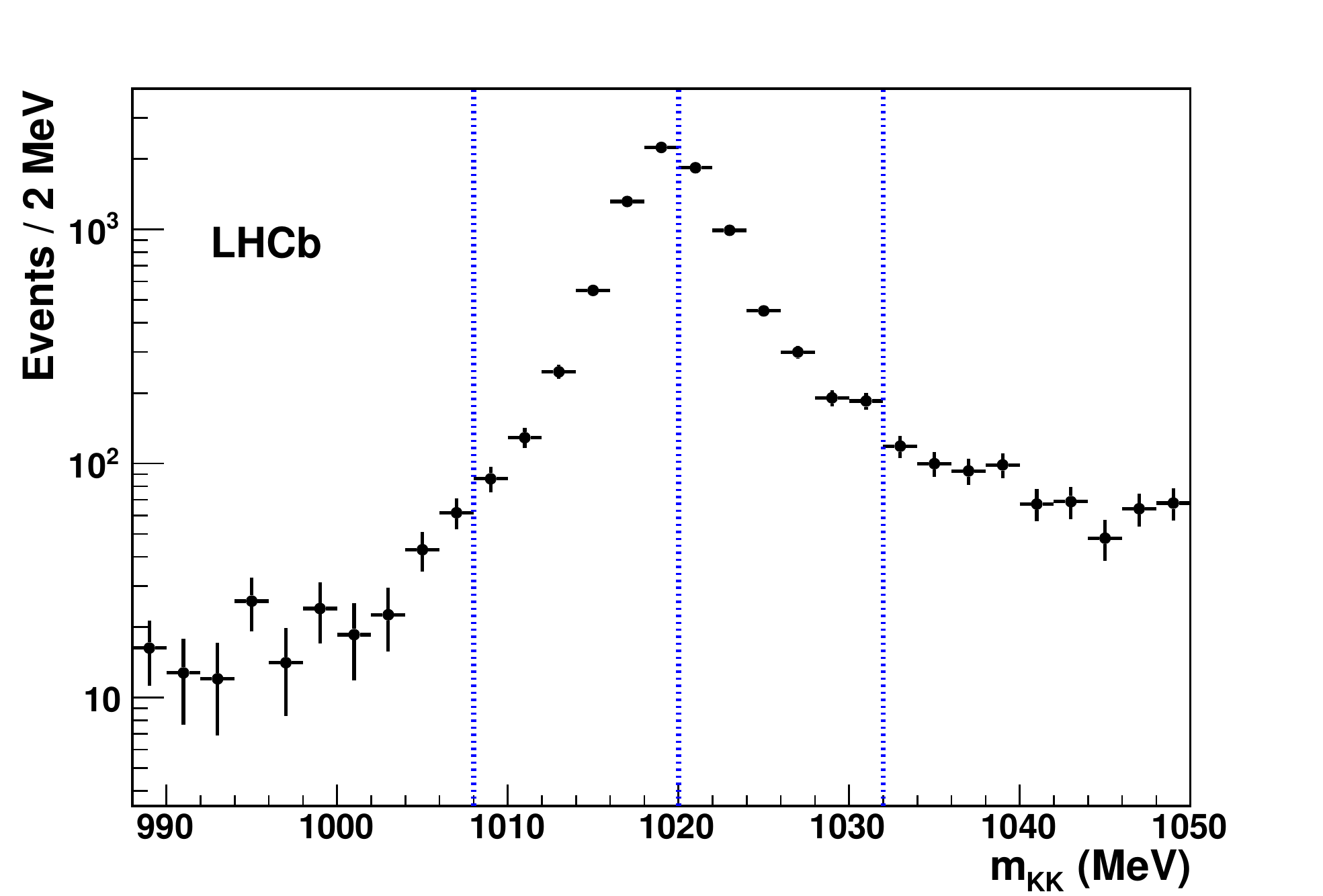}
  }
  \caption{ Background subtracted \KpKm\ invariant  mass distribution for $\BsToJPsiKK$ candidates. The vertical dotted lines separate the
   four intervals.}
\label{fig:mKK}
\end{figure}

This analysis  uses  the sWeight technique~\cite{Pivk:2004ty} for background subtraction.
The signal weight, denoted by $W_{\rm s}(m_{J/\psi KK})$, 
is obtained using $m_{J/\psi KK}$ as the discriminating variable. 
The correlations between   $m_{J/\psi KK}$ and other variables  used
in the analysis, including $m_{KK}$, \mbox{decay} time $t$ and the angular variables $\Omega$
 defined in Ref.~\cite{LHCb:2011aa},  are found to be
negligible for both the signal and background components in the data. 
Figure~\ref{fig:mKK} shows the $m_{KK}$ distribution where the  background is subtracted statistically using the sWeight technique.
The range of $m_{KK}$ is divided into four intervals: 988--1008, 1008--1020,
1020--1032  and 1032--1050 MeV.
Table~\ref{tab:binning} gives the number of \Bs\ signal and background  candidates in 
each interval.

\begin{table}[htbp]
\caption{Numbers of signal and background events in the $m_{J/\psi KK}$ range of 5200--5550 MeV
and statistical power per signal event in four intervals of $m_{KK}$.\\}
\begin{center}
\begin{tabular}{c|c|c|c|c}
\hline
 $k$ & $m_{KK}$ interval (MeV) &  $N_{{\rm sig};k}$& $N_{{\rm bkg};k}$  & $W_{{\rm p};k}$   \\\hline
 1 & $\phantom{0}$988--1008   &  $\phantom{0}251 \pm 21$ &   $1675 \pm 43$  & 0.700  \\
 2 & 1008--1020   &   $4569 \pm 70$ &  $2002 \pm 49$  & 0.952  \\
 3 & 1020--1032   &   $3952 \pm 66$ &  $2244 \pm 51$  & 0.938 \\
 4 & 1032--1050   &   $\phantom{0} 726 \pm 34$ &  $3442 \pm 62$ & 0.764 \\ \hline
\end{tabular}
\end{center}
\label{tab:binning}
\end{table}

\begin{figure}[t]
\vfill
   \includegraphics[angle=0,width=0.5\textwidth]{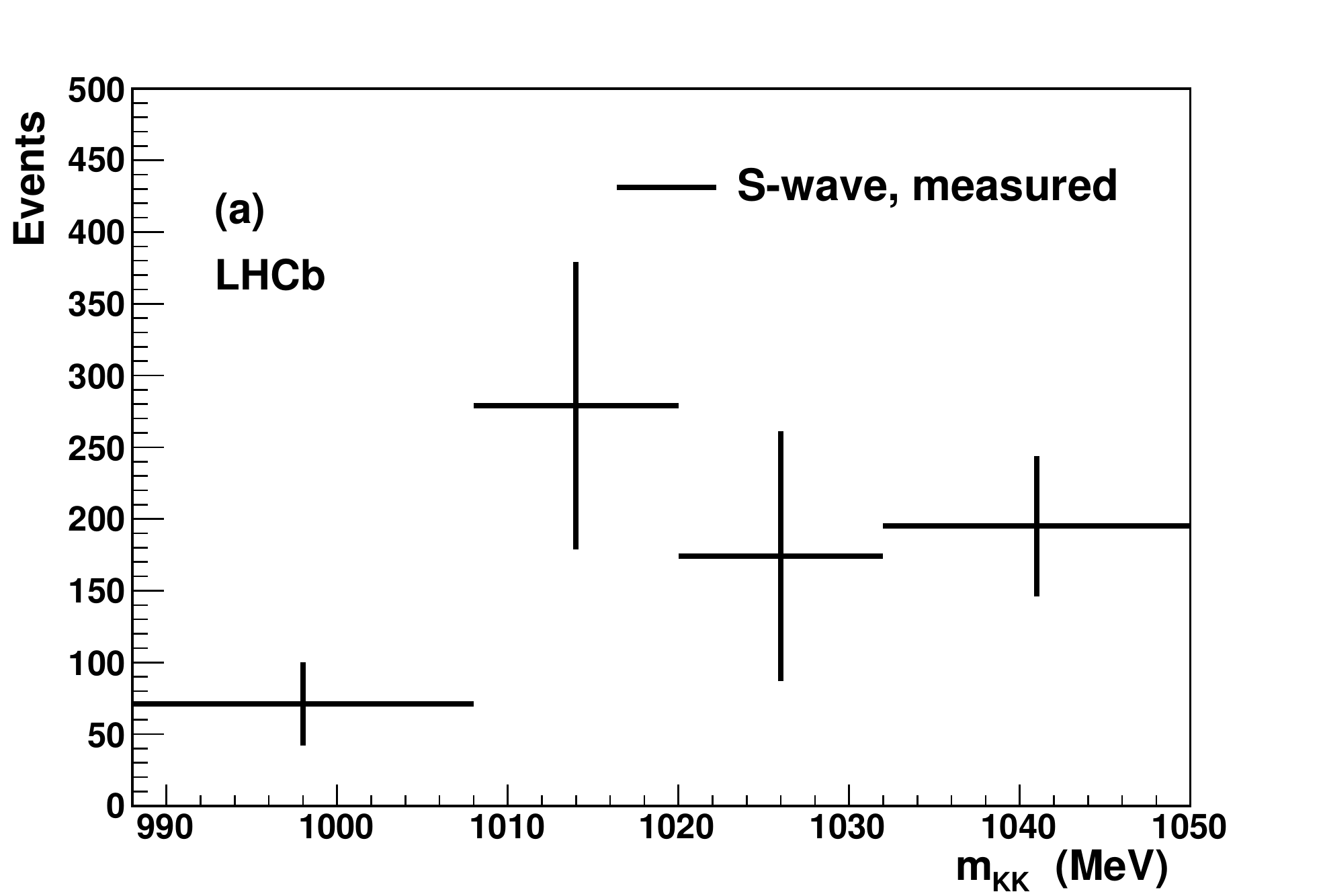}
   \includegraphics[angle=0,width=0.5\textwidth]{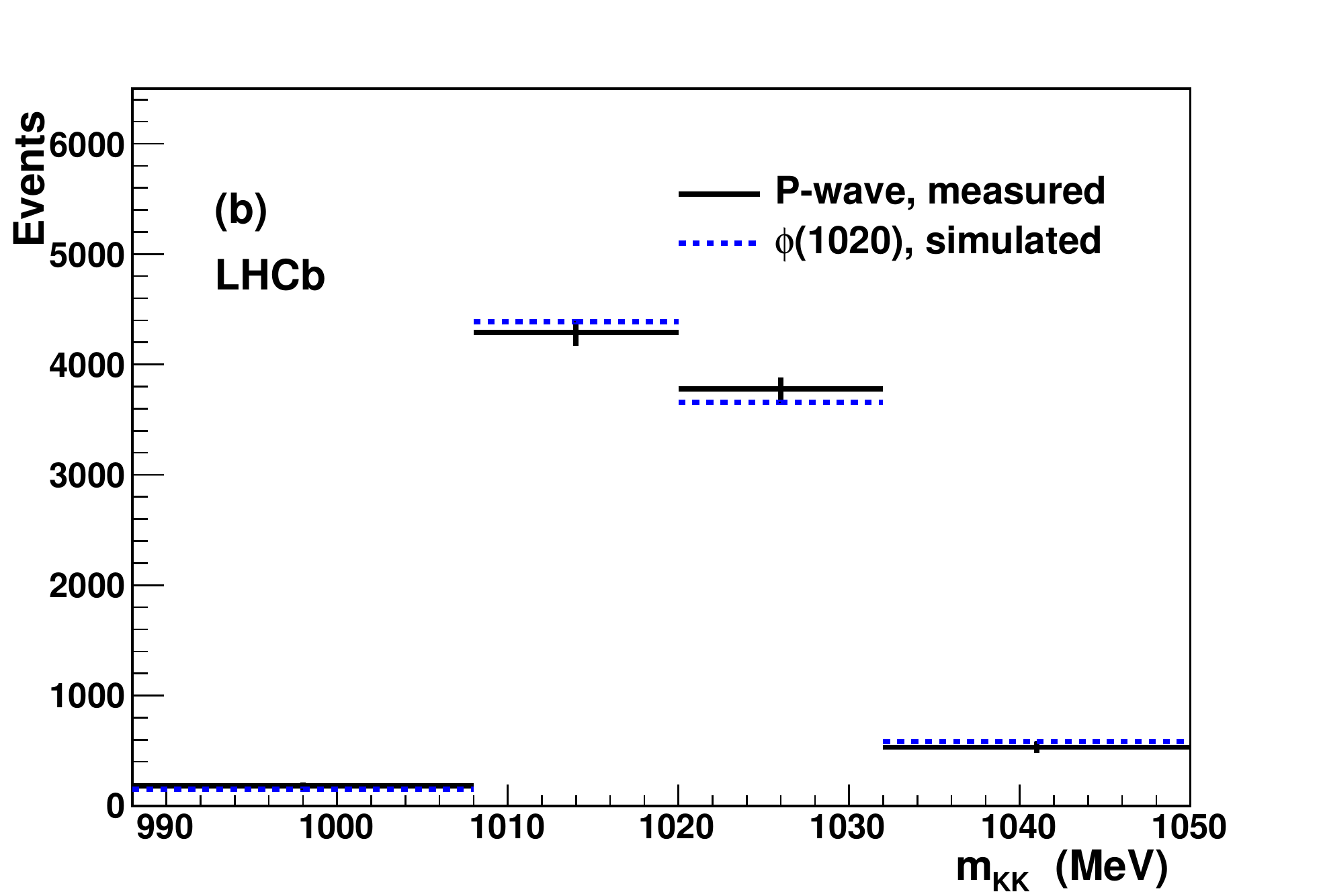}
  \caption{ Distribution of 
  (a) \KpKm\ S-wave signal events, and (b) \KpKm\ P-wave signal events,
   both in four invariant mass intervals. In (b), the distribution of simulated \BsToJPsiPhi\  events
        in the four intervals assuming the same total number of P-wave events is also shown (dashed lines).
   Note the interference between the \KpKm\ S-wave and P-wave amplitudes integrated over
   the angular variables has vanishing contribution in these distributions.  
}
  \label{fig:mkk_SPwave}
\end{figure}

\begin{table}[t]
\caption{Results from a simultaneous fit of the four intervals of $m_{KK}$,
where the uncertainties are statistical only. Only parameters which are needed for the ambiguity 
resolution are shown.\\
}
\begin{center}
\begin{tabular}{|l|c|c|}
\hline
           Parameter            &  Solution I  & Solution II        \\ \hline \hline
            $\phi_s$   (rad)         &      0.167 $\pm$     0.175   &  \phantom{$-$}2.975 $\pm$     0.175               \\
      $\Delta\Gamma$  ($\ps ^{-1}$)          &       0.120 $\pm$    0.028  &  $-0.120$ $\pm$    0.028            \\
           ${F_{\mathrm{S};1}}$            &       0.283 $\pm$    0.113        &  \phantom{$-$}0.283 $\pm$    0.113                      \\
           ${F_{\mathrm{S};2}}$            &       0.061 $\pm$    0.022    &  \phantom{$-$}0.061 $\pm$    0.022                        \\
           ${F_{\mathrm{S};3}}$            &       0.044 $\pm$    0.022   &  \phantom{$-$}0.044 $\pm$    0.022                       \\
           ${F_{\mathrm{S};4}}$            &       0.269 $\pm$    0.067    & \phantom{$-$}0.269 $\pm$    0.067                     \\
           $\delta_{\mathrm{S}\perp;1}$   (rad)         &   \phantom{$-$}$2.68 \,\, _{-\, 0.42}^{+\, 0.35}$    &  $0.46 \,\, ^{+\, 0.42}_{-\, 0.35}$  \\
           $\delta_{\mathrm{S}\perp;2}$   (rad)         &   \phantom{$-$}$0.22 \,\, _{-\, 0.13}^{+\, 0.15}$    &  $2.92 \,\, ^{+\, 0.13}_{-\, 0.15}$   \\
           $\delta_{\mathrm{S}\perp;3}$   (rad)         &   $-0.11 \,\, _{-\, 0.18}^{+\, 0.16}$    &  $3.25 \,\, ^{+\, 0.18}_{-\, 0.16}$    \\
           $\delta_{\mathrm{S}\perp;4}$   (rad)         &   $-0.97 \,\, _{-\, 0.43}^{+\, 0.28}$    &  $4.11 \,\, ^{+\, 0.43}_{-\, 0.28}$      \\
            \hline
\end{tabular}
\end{center}
\label{tab:results-4bin}
\end{table}

\begin{figure}[htbp]
  \centerline{
   \includegraphics[angle=0,width=0.5\textwidth]{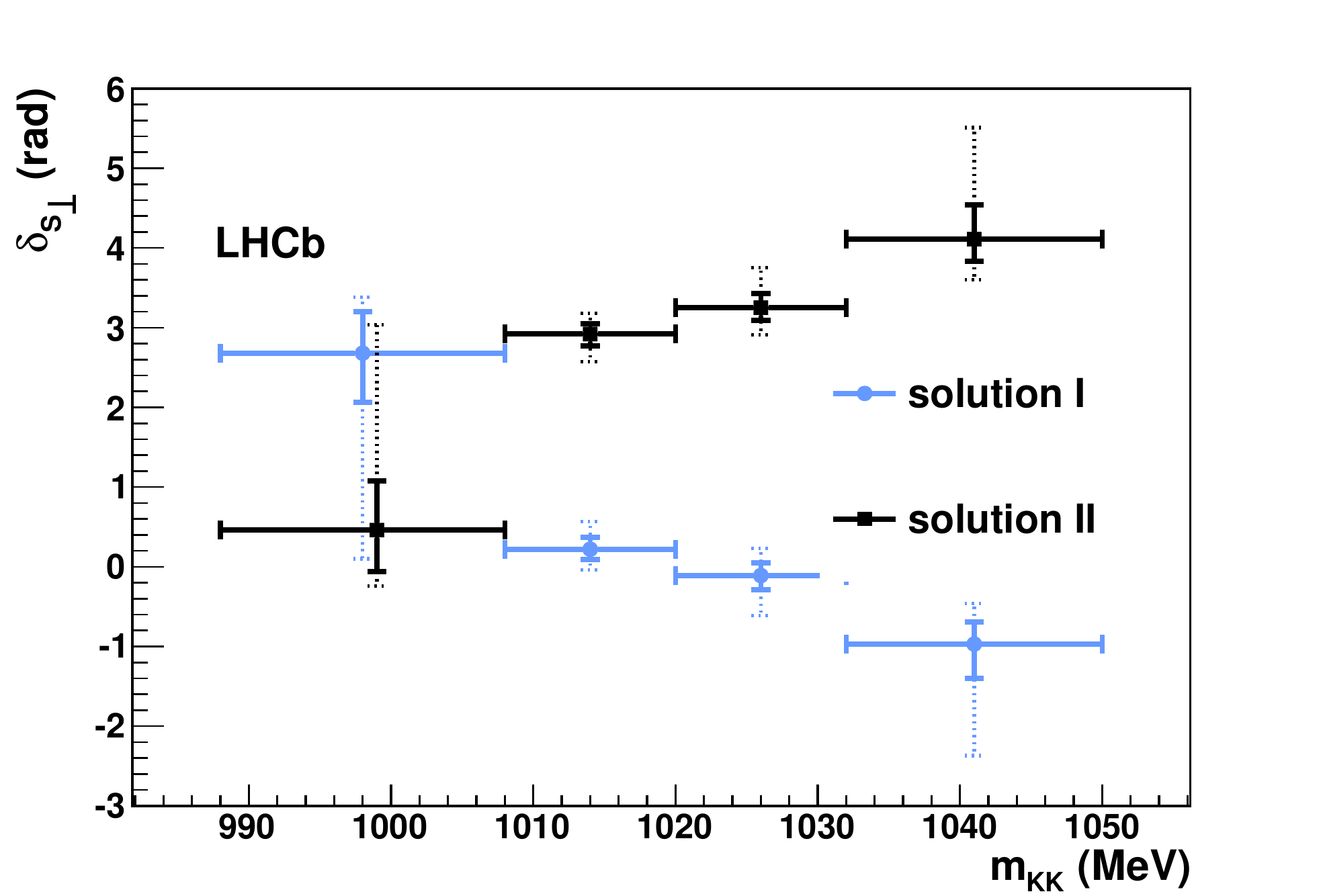}
  }
  \caption{ Measured phase differences between S-wave and perpendicular P-wave amplitudes
  in four intervals of $m_{KK}$ for solution I (full blue  circles) and solution II (full black squares).
   The asymmetric error bars correspond to  $\Delta\ln L =-0.5$ (solid lines)   and  $\Delta\ln L =-2$ (dash-dotted lines).
}
\label{fig:deltaS_mKK}
\end{figure}


In this analysis we perform an unbinned maximum likeli\-hood fit to the  data
using the sFit method~\cite{2009arXiv0905.0724X},  an extension of the sWeight technique,  
that simplifies  fitting in the presence of background.
In this method, it is only necessary to model the signal PDF,  as background is cancelled statistically using the signal weights.

The parameters of the  $\BsToJPsiKK$ decay time distribution are estimated from
a simultaneous fit to the four intervals of $m_{KK}$
by maximizing the log-likelihood function
\begin{eqnarray}
   \ln L && ({\bf \Theta_P, \Theta_\mathrm{S}} )  =   \sum_{k=1}^4   W_{{\rm p};k}    \sum_{i=1}^{N_k}  W_{\rm s} (m_{J/\psi KK ;i}) \times \nonumber  \\
& & \ln P_{\rm sig}(t_i, \Omega_i, q_i, \omega_i; {\bf \Theta_P}, {\bf \Theta_\mathrm{S}}  ),\,  \nonumber 
\label{eq:lnL_4bin}
\end{eqnarray}
where $N_k = N_{{\rm sig};k}+N_{{\rm bkg};k}$ is the number of candidates in the $m_{J/\psi KK}$ range of 5200--5550 MeV for the $k$th interval.
$\bf \Theta_P$ represents the physics   parameters  independent of $m_{KK}$, including \phis, \DGs\ and
the magnitudes and phases of the P-wave amplitudes. 
Note that the  P-wave amplitudes for different polarizations share the same dependence on $m_{KK}$. 
$\bf \Theta_\mathrm{S}$  denotes the  values of the  $m_{KK}$-dependent parameters
averaged over each  interval, 
namely the  average fraction of S-wave contribution for the $k$th interval,
$F_{\mathrm{S};k}$,  and the average phase difference between the S-wave amplitude and the perpendicular P-wave amplitude
for the $k$th interval, $\delta_{\mathrm{S}\perp ;k}$.
$P_{\rm sig}$ is the  signal PDF  of the decay time $t$, angular variables $\Omega$,   initial flavour tag $q$
and the  mistag probability $\omega$. It is based on the theoretical differential
decay rates~\cite{Xie:2009fs} and includes experimental effects such as decay time resolution and acceptance,
angular acceptance and imperfect identification of the  initial flavour of the \Bs\ particle, 
as described in Ref.~\cite{LHCb:2011aa}. The factors $W_{{\rm p};k}$
account for loss of statistical precision in parameter estimation due to  background dilution
 and are necessary to obtain the correct error coverage.
Their values are given in Table~\ref{tab:binning}.

The fit results for \phis, \DGs, $F_{\mathrm{S};k}$ and $\delta_{\mathrm{S}\perp ;k}$
are given in Table~\ref{tab:results-4bin}.
 Figure~\ref{fig:mkk_SPwave} shows the estimated \KpKm\ S-wave and P-wave contributions in the four $m_{KK}$ intervals.
The shape of the measured P-wave $m_{KK}$ distribution is  in good agreement with
 that of  \BsToJPsiPhi\  events simulated using  a spin-1 relativistic Breit-Wigner 
function for the \particle{\phi (1020)} amplitude.
In Fig.~\ref{fig:deltaS_mKK},
the phase difference between the S-wave and the perpendicular P-wave amplitude
 is plotted   in four  $m_{KK}$  intervals  for solution I and solution II.

Figure~\ref{fig:deltaS_mKK} shows a clear decreasing trend 
of the  phase difference between the S-wave and P-wave amplitudes
in the  \particle{\phi (1020)} mass region
for solution I, as expected for the physical solution. To estimate the significance of the result, we perform 
an unbinned maximum likelihood fit to the data  by parameterizing the
  phase difference $\delta_{{\mathrm{S}}\perp;k}$ as a linear function of 
the average $m_{KK}$ value in the $k$th interval. This
leads to a slope of $-0.050 _ {-0.020} ^{+0.013}$ rad/MeV for solution I
and the opposite sign  for solution II,
where the uncertainties are statistical only. 
The difference of the  $\ln L$ value between this fit 
and a fit in which the slope is fixed to be  zero is 11.0.
Hence, the negative trend of solution I has a significance of 4.7 standard deviations.
Therefore, we conclude that solution I, which has  $\DGs>0$,  is the physical solution.
The trend of  solution I is also qualitatively consistent with that  of the phase difference between the $\KpKm$
S-wave and P-wave amplitudes  versus $m_{KK}$  
measured in the decay $D_s^+ \to K^+ K^- \pi^+$ by the BaBar collaboration~\cite{delAmoSanchez:2010yp}.

Several possible sources of systematic uncertainty on the phase variation versus $m_{KK}$ have been considered.   
A possible background  from decays with similar final states such as $\BdToJPsiKst$ could have a small effect.
From  simulation, the contamination to the signal
from such decays is estimated to be $1.1\%$
in the $m_{KK}$ range of 988--1050 MeV.
We add a  $2.2\%$  contribution of simulated \BdToJPsiKst\ events to the data
and repeat the analysis. The largest observed change is a  shift of $\delta_{\mathrm{S}\perp;4}$  by 
$0.06$ rad, which is only 20$\%$ of its statistical uncertainty and has negligible effect on the
slope of $\delta_{\mathrm{S}\perp}$   versus $m_{KK}$.
The effect of neglecting the variation of the values of $F_{\mathrm{S}}$ and $\delta_{\mathrm{S}\perp}$ 
in each $m_{KK}$ interval 
is determined to change  the significance of the negative trend of solution I
by less than 0.1 standard deviations.
We also repeat the analysis for different $m_{KK}$ ranges,   different
ways of dividing the $m_{KK}$ range, or different shapes of the signal and background $m_{J/\psi KK}$
distributions. The  significance of the negative trend of solution I is not affected. 
To measure precisely the S-wave line shape and  determine its resonance structure, more data are needed.
However, the results presented here do not depend on such detailed knowledge.

In conclusion the analysis of the strong interaction phase shift resolves the ambiguity between solution I and 
solution II.
Values of $\phis$ close to zero and  positive  \DGs\ are preferred.
It follows that in the \Bs\ system,
the mass eigenstate that is almost $\CP$ even is lighter and \mbox{decays} faster than the state 
that is almost $\CP$ odd. 
This is in  agreement with the Standard Model expectations (e.g.,~\cite{Lenz:2010gu}). 
It is also interesting to note that  this situation is similar to that in the neutral kaon system.

\section*{Acknowledgements}

\noindent We express our gratitude to our colleagues in the CERN accelerator
departments for the excellent performance of the LHC. We thank the
technical and administrative staff at CERN and at the LHCb institutes,
and acknowledge support from the National Agencies: CAPES, CNPq,
FAPERJ and FINEP (Brazil); CERN; NSFC (China); CNRS/IN2P3 (France);
BMBF, DFG, HGF and MPG (Germany); SFI (Ireland); INFN (Italy); FOM and
NWO (The Netherlands); SCSR (Poland); ANCS (Romania); MinES of Russia and
Rosatom (Russia); MICINN, XuntaGal and GENCAT (Spain); SNSF and SER
(Switzerland); NAS Ukraine (Ukraine); STFC (United Kingdom); NSF
(USA). We also acknowledge the support received from the ERC under FP7
and the Region Auvergne.


\bibliographystyle{LHCb}
\bibliography{main}

\ifx\mcitethebibliography\mciteundefinedmacro
\PackageError{LHCb.bst}{mciteplus.sty has not been loaded}
{This bibstyle requires the use of the mciteplus package.}\fi
\providecommand{\href}[2]{#2}
\begin{mcitethebibliography}{10}
\mciteSetBstSublistMode{n}
\mciteSetBstMaxWidthForm{subitem}{\alph{mcitesubitemcount})}
\mciteSetBstSublistLabelBeginEnd{\mcitemaxwidthsubitemform\space}
{\relax}{\relax}

\bibitem{Abazov:2011ry}
\Dzero{} collaboration, V.~M. Abazov {\em et~al.},
  \ifthenelse{\boolean{articletitles}}{{\it {Measurement of the \CP\-violating
  phase $\phi_s^{J/\psi \phi}$ using the flavor-tagged decay \BsToJPsiPhi\ in 8
  \invfb\ of $p {\bar p}$ collisions}},
  }{}\href{http://dx.doi.org/10.1103/PhysRevD.85.032006}{Phys.\ Rev.\  {\bf
  D85} (2012) 032006}, \href{http://xxx.lanl.gov/abs/1109.3166}{{\tt
  arXiv:1109.3166}}\relax
\mciteBstWouldAddEndPuncttrue
\mciteSetBstMidEndSepPunct{\mcitedefaultmidpunct}
{\mcitedefaultendpunct}{\mcitedefaultseppunct}\relax
\EndOfBibitem
\bibitem{CDF:2011af}
CDF collaboration, T.~Aaltonen {\em et~al.},
  \ifthenelse{\boolean{articletitles}}{{\it {Measurement of the \CP\-violating
  phase $\beta_s$ in \BsToJPsiPhi\ decays with the CDF II Detector}},
  }{}\href{http://xxx.lanl.gov/abs/1112.1726}{{\tt arXiv:1112.1726}}\relax
\mciteBstWouldAddEndPuncttrue
\mciteSetBstMidEndSepPunct{\mcitedefaultmidpunct}
{\mcitedefaultendpunct}{\mcitedefaultseppunct}\relax
\EndOfBibitem
\bibitem{LHCb:2011aa}
LHCb collaboration, R.~Aaij {\em et~al.},
  \ifthenelse{\boolean{articletitles}}{{\it {Measurement of the CP-violating
  phase \phis\ in the decay $\BsToJPsiPhi$}},
  }{}\href{http://dx.doi.org/10.1103/PhysRevLett.108.101803}{Phys.\ Rev.\
  Lett.\  {\bf 108} (2012) 101803},
  \href{http://xxx.lanl.gov/abs/1112.3183}{{\tt arXiv:1112.3183}}\relax
\mciteBstWouldAddEndPuncttrue
\mciteSetBstMidEndSepPunct{\mcitedefaultmidpunct}
{\mcitedefaultendpunct}{\mcitedefaultseppunct}\relax
\EndOfBibitem
\bibitem{Dunietz:2000cr}
I.~Dunietz, R.~Fleischer, and U.~Nierste,
  \ifthenelse{\boolean{articletitles}}{{\it {In pursuit of new physics with
  $\Bs$ decays}}, }{}\href{http://dx.doi.org/10.1103/PhysRevD.63.114015}{Phys.\
  Rev.\  {\bf D63} (2001) 114015},
  \href{http://xxx.lanl.gov/abs/hep-ph/0012219}{{\tt
  arXiv:hep-ph/0012219}}\relax
\mciteBstWouldAddEndPuncttrue
\mciteSetBstMidEndSepPunct{\mcitedefaultmidpunct}
{\mcitedefaultendpunct}{\mcitedefaultseppunct}\relax
\EndOfBibitem
\bibitem{Nandi:2008rg}
S.~Nandi and U.~Nierste, \ifthenelse{\boolean{articletitles}}{{\it {Resolving
  the sign ambiguity in \DGs\ with $\Bs \rightarrow \D_s K$}},
  }{}\href{http://dx.doi.org/10.1103/PhysRevD.77.054010}{Phys.\ Rev.\  {\bf
  D77} (2008) 054010}, \href{http://xxx.lanl.gov/abs/0801.0143}{{\tt
  arXiv:0801.0143}}\relax
\mciteBstWouldAddEndPuncttrue
\mciteSetBstMidEndSepPunct{\mcitedefaultmidpunct}
{\mcitedefaultendpunct}{\mcitedefaultseppunct}\relax
\EndOfBibitem
\bibitem{Xie:2009fs}
Y.~Xie, P.~Clarke, G.~Cowan, and F.~Muheim,
  \ifthenelse{\boolean{articletitles}}{{\it {Determination of $2\beta_s$ in
  \BsToJPsiKK\ decays in the presence of a \KpKm\ S-wave contribution}},
  }{}\href{http://dx.doi.org/10.1088/1126-6708/2009/09/074}{JHEP {\bf 09}
  (2009) 074}, \href{http://xxx.lanl.gov/abs/0908.3627}{{\tt
  arXiv:0908.3627}}\relax
\mciteBstWouldAddEndPuncttrue
\mciteSetBstMidEndSepPunct{\mcitedefaultmidpunct}
{\mcitedefaultendpunct}{\mcitedefaultseppunct}\relax
\EndOfBibitem
\bibitem{Aubert:2004cp}
BaBar collaboration, B.~Aubert {\em et~al.},
  \ifthenelse{\boolean{articletitles}}{{\it {Ambiguity-free measurement of
  $\cos2\beta$: time-integrated and time-dependent angular analyses of $B
  \rightarrow J/\psi K \pi$}},
  }{}\href{http://dx.doi.org/10.1103/PhysRevD.71.032005}{Phys.\ Rev.\  {\bf
  D71} (2005) 032005}, \href{http://xxx.lanl.gov/abs/hep-ex/0411016}{{\tt
  arXiv:hep-ex/0411016}}\relax
\mciteBstWouldAddEndPuncttrue
\mciteSetBstMidEndSepPunct{\mcitedefaultmidpunct}
{\mcitedefaultendpunct}{\mcitedefaultseppunct}\relax
\EndOfBibitem
\bibitem{Alves:2008zz}
LHCb collaboration, J.~Alves, A.~Augusto {\em et~al.},
  \ifthenelse{\boolean{articletitles}}{{\it {The LHCb Detector at the LHC}},
  }{}\href{http://dx.doi.org/10.1088/1748-0221/3/08/S08005}{JINST {\bf 3}
  (2008) S08005}\relax
\mciteBstWouldAddEndPuncttrue
\mciteSetBstMidEndSepPunct{\mcitedefaultmidpunct}
{\mcitedefaultendpunct}{\mcitedefaultseppunct}\relax
\EndOfBibitem
\bibitem{Abazov:2011hv}
\Dzero{} collaboration, V.~M. Abazov {\em et~al.},
  \ifthenelse{\boolean{articletitles}}{{\it {Measurement of the relative
  branching ratio of $\Bs \to \jpsi f_0$ to $\BsToJPsiPhi$}},
  }{}\href{http://dx.doi.org/10.1103/PhysRevD.85.011103}{Phys.\ Rev.\  {\bf
  D85} (2012) 011103}, \href{http://xxx.lanl.gov/abs/1110.4272}{{\tt
  arXiv:1110.4272}}\relax
\mciteBstWouldAddEndPuncttrue
\mciteSetBstMidEndSepPunct{\mcitedefaultmidpunct}
{\mcitedefaultendpunct}{\mcitedefaultseppunct}\relax
\EndOfBibitem
\bibitem{Stone:2008ak}
S.~Stone and L.~Zhang, \ifthenelse{\boolean{articletitles}}{{\it {S-waves and
  the measurement of \CP\ violating phases in \Bs\ decays}},
  }{}\href{http://dx.doi.org/10.1103/PhysRevD.79.074024}{Phys.\ Rev.\  {\bf
  D79} (2009) 074024}, \href{http://xxx.lanl.gov/abs/0812.2832}{{\tt
  arXiv:0812.2832}}\relax
\mciteBstWouldAddEndPuncttrue
\mciteSetBstMidEndSepPunct{\mcitedefaultmidpunct}
{\mcitedefaultendpunct}{\mcitedefaultseppunct}\relax
\EndOfBibitem
\bibitem{Pivk:2004ty}
M.~Pivk and F.~R. Le~Diberder, \ifthenelse{\boolean{articletitles}}{{\it
  {sPlot: a statistical tool to unfold data distributions}},
  }{}\href{http://dx.doi.org/10.1016/j.nima.2005.08.106}{Nucl.\ Instrum.\
  Meth.\  {\bf A555} (2005) 356},
  \href{http://xxx.lanl.gov/abs/physics/0402083}{{\tt
  arXiv:physics/0402083}}\relax
\mciteBstWouldAddEndPuncttrue
\mciteSetBstMidEndSepPunct{\mcitedefaultmidpunct}
{\mcitedefaultendpunct}{\mcitedefaultseppunct}\relax
\EndOfBibitem
\bibitem{2009arXiv0905.0724X}
Y.~{Xie}, \ifthenelse{\boolean{articletitles}}{{\it {sFit: a method for
  background subtraction in maximum likelihood fit}},
  }{}\href{http://xxx.lanl.gov/abs/0905.0724}{{\tt arXiv:0905.0724}}\relax
\mciteBstWouldAddEndPuncttrue
\mciteSetBstMidEndSepPunct{\mcitedefaultmidpunct}
{\mcitedefaultendpunct}{\mcitedefaultseppunct}\relax
\EndOfBibitem
\bibitem{delAmoSanchez:2010yp}
BaBar collaboration, P.~del Amo~Sanchez {\em et~al.},
  \ifthenelse{\boolean{articletitles}}{{\it {Dalitz plot analysis of $\Ds
  \rightarrow \KpKm \pip$}},
  }{}\href{http://dx.doi.org/10.1103/PhysRevD.83.052001}{Phys.\ Rev.\  {\bf
  D83} (2011) 052001}, \href{http://xxx.lanl.gov/abs/1011.4190}{{\tt
  arXiv:1011.4190}}\relax
\mciteBstWouldAddEndPuncttrue
\mciteSetBstMidEndSepPunct{\mcitedefaultmidpunct}
{\mcitedefaultendpunct}{\mcitedefaultseppunct}\relax
\EndOfBibitem
\bibitem{Lenz:2010gu}
A.~Lenz {\em et~al.}, \ifthenelse{\boolean{articletitles}}{{\it {Anatomy of new
  physics in $\BBbar$ mixing}},
  }{}\href{http://dx.doi.org/10.1103/PhysRevD.83.036004}{Phys.\ Rev.\  {\bf
  D83} (2011) 036004}, \href{http://xxx.lanl.gov/abs/1008.1593}{{\tt
  arXiv:1008.1593}}\relax
\mciteBstWouldAddEndPuncttrue
\mciteSetBstMidEndSepPunct{\mcitedefaultmidpunct}
{\mcitedefaultendpunct}{\mcitedefaultseppunct}\relax
\EndOfBibitem
\end{mcitethebibliography}

\end{document}